\documentclass[intlimits,twoside,a4paper]{article}

\usepackage[cp1251]{inputenc}

\usepackage{amsmath,amssymb}
\usepackage{graphicx}

\usepackage[eqsecnum]{cmpj3}

\usepackage{bm}

\issue{2019}{22}{4}{43301}
\doinumber{10.5488/CMP.22.43301}

\title[Synchrotron-based UV resonance Raman scattering for investigating ionic liquid-water solutions]%
{Synchrotron-based UV resonance Raman scattering for investigating ionic liquid-water solutions%
 \protect\\ [0.9ex] }
\author[C. Bottari \textsl{et al.}]{C. Bottari\refaddr{label1,label2},
        B. Rossi\refaddr{label1}, A. Mele\refaddr{label3,label4}, A. Damin\refaddr{label5}, S. Bordiga\refaddr{label5}, M. Musso\refaddr{label6}, A. Gessini\refaddr{label1}, C.~Masciovecchio\refaddr{label1}}
\addresses{
\addr{label1} Elettra-Sincrotrone Trieste, S.S. 114 km 163.5, Basovizza, 34149 Trieste, Italy
\addr{label2} Department of Physics, University of Trieste, Via A. Valerio, 2, 34127 Trieste, Italy
\addr{label3} Department of Chemistry, Materials and Chemical Engineering ``G. Natta'', Politecnico di Milano, Piazza L. da Vinci, 32, 20133 Milano, Italy
\addr{label4} CNR-ICRM, Via L. Mancinelli, 7, 20131 Milano, Italy 
\addr{label5} Department of Chemistry, NIS Centre and INSTM Reference Centre University of Turin, Via G. Quarello, 15, 10135 Turin, Italy
\addr{label6} Department of Chemistry and Physics of Materials, University of Salzburg, Salzburg, Austria
}

\date{Received	May 2, 2019}

\begin{document}

\maketitle

\begin{abstract}
This work shows that bulk ionic liquids (ILs) and their water solution can be conveniently investigated by synchrotron-based UV resonance Raman (UVRR) spectroscopy. The main advantages of this technique for the investigation of the local structure and intermolecular interactions in imidazolium-based ILs are presented and discussed. The unique tunability of synchrotron source allows one to selectively enhance in the Raman spectra the vibrational signals arising from the imidazolium ring. Such signals showed good sensitivity to the modifications induced in the local structure of ILs by i) the change of  anion and ii) the progressively longer alkyl chain substitution on the imidazolium ring. Moreover, some UVRR signals are specifically informative on the effect induced by addition of water on the strength of cation-anion H-bonds in IL-water solutions. All of these results corroborate the potentiality of UVRR to retrieve information on the intermolecular interactions in IL-water solutions, besides the counterpart obtained by employing on these systems the spontaneous Raman scattering technique.
\pacs 39.30.+w, 82.90.+j
\end{abstract}

\section{Introduction}
Ionic liquids (ILs) belong to a broad class of ionic compounds that, differently from conventional salts, are usually liquid at T<100 \textcelsius. They are characterized by the vanishing vapour pressure, good thermal stability, high ion density and ionic conductivity \cite{1,2}. Several applications of ILs have been explored, including their use in organic synthesis \cite{3}, electrochemical devices \cite{4}, photochemical cells~\cite{5,6} and catalysis \cite{7}. Thanks to a large variety of available ions, the physico-chemical properties of ILs can be modulated by careful selection of both cation and anion with specific characteristics for tailored applications \cite{8}. A more convenient strategy for an efficient tuning of the performances of ILs consists in mixing ILs with other ionic or molecular liquids, such as e.g., water \cite{9}. Due to the strong hygroscopic nature of most ILs, the presence of trace amounts of water was originally considered problematic for maintaining the peculiar properties of ILs. Nowadays, it is observed that addition of water to ILs allows one to improve some of their properties and performances towards specific applications. For instance, water incorporation in ILs implies an increase of the self-diffusion coefficient of cations and anions due to a general decrease in the viscosity \cite{10}, leading, to a certain extent, to a larger ionic conductivity \cite{11}. In the biological field, water represents an ideal partner for ILs and several recent studies reported on the capability of IL/water solutions to increase the enzyme activity \cite{12} and improve the stabilization of proteins \cite{13,14,15}. Since water has such a high impact on the chemical-physical characteristics of ILs, both as  a contaminant or as co-solvent, a detailed knowledge of the intermolecular interactions taking place in IL-water solutions is a crucial step for understanding and predicting the range of properties of ILs \cite{16}. This appears particularly challenging in the case of ILs, where the scenario is complicated due to the fact that intermolecular interactions can be strongly affected by several possible cation-anion combinations.
Several methods have been used to get insights into the molecular dynamics of pure ILs and mixed with water, including vibrational spectroscopy \cite{17}. Such techniques makes it possible to investigate the extension and the strength of inter- and intra-molecular interactions in molecular liquids by probing to the vibrational motions of the system. For instance, collective vibrations of pure liquid water have been successfully investigated by using low-frequency Raman spectroscopy through the comparison between experimental data and molecular dynamics simulations \cite{18}. Apart from the niche applications of low frequency Raman spectroscopy, the usual spectral range between $400$--$3800$ cm$^{-1}$ can be exploited to monitor the variation of the chemical environment around the oscillators. Cammarata \textsl{et al.}~\cite{19} used IR spectroscopy to correlate the blue-shift of the asymmetric stretching band of water with the relative strength of the interaction of H-bonding between water molecules and different anions species. Moreover, such H-bond strength has been correlated to the degree of hydrophobicity \cite{20} and the polarity \cite{21} of ILs. In another paper, Andanson \textsl{et al.}~\cite{22} used IR spectroscopy for understanding, at a nanoscopic scale, the mixing behaviour of ILs. The work demonstrated that vibrational techniques are suitable for the detection and quantification of the coexistence of different species of water clusters into domains of ILs. The local organization of water molecules in IL-water solutions has been investigated by using both Raman and IR spectroscopy \cite{23,24}. The conformational stability of imidazolium-based ILs in the presence of water has been discussed by Hatano \textsl{et al.} \cite{25} by exploiting the information extracted from the Raman spectra of these solutions. Finally, Raman and IR spectroscopy were successfully employed for investigating the effect of water on the local structure and the phase behavior of protic ILs \cite{26}, giving evidence of the strong complementarity of the two techniques in describing different water-IL interactions. 

Herein, we provide the experimental evidence of the potentiality of synchrotron based-UV Resonance Raman (UVRR) technique to probe the structural organization and the intermolecular interactions in imidazolium-based ILs such as pure liquids and in water solution. UVRR exhibits several advantages with respect to the conventional spontaneous Raman technique: i) a significant increment of the detection limit that allows one to investigate the vibrational modes of ILs also in very high diluted conditions and ii) a selective strong enhancement in the UVRR spectra of ILs of the Raman cross-section of the vibrations involving the imidazolium ring. This latter condition occurs in particular when the excitation wavelength of UVRR spectra approaches the $\pi$-$\pi$\textsuperscript{*} transitions in the deep UV range \cite{27,28,29,30}. In this sense, the availability of a tuneable UV synchrotron radiation (SR) source allows one to finely choose the resonance energy of excitation wavelength in order to maximize the intensity of the Raman peaks associated with the cations of ILs. These signals are sensitive spectroscopic markers of local rearrangements occurring in ILs and of the interactions between molecular domains of ILs and water molecules.

\section{Experimental Method}
\subsection*{Sample prepatation}
\begin{figure}[!t]
\centerline{\includegraphics[width=0.50\textwidth]{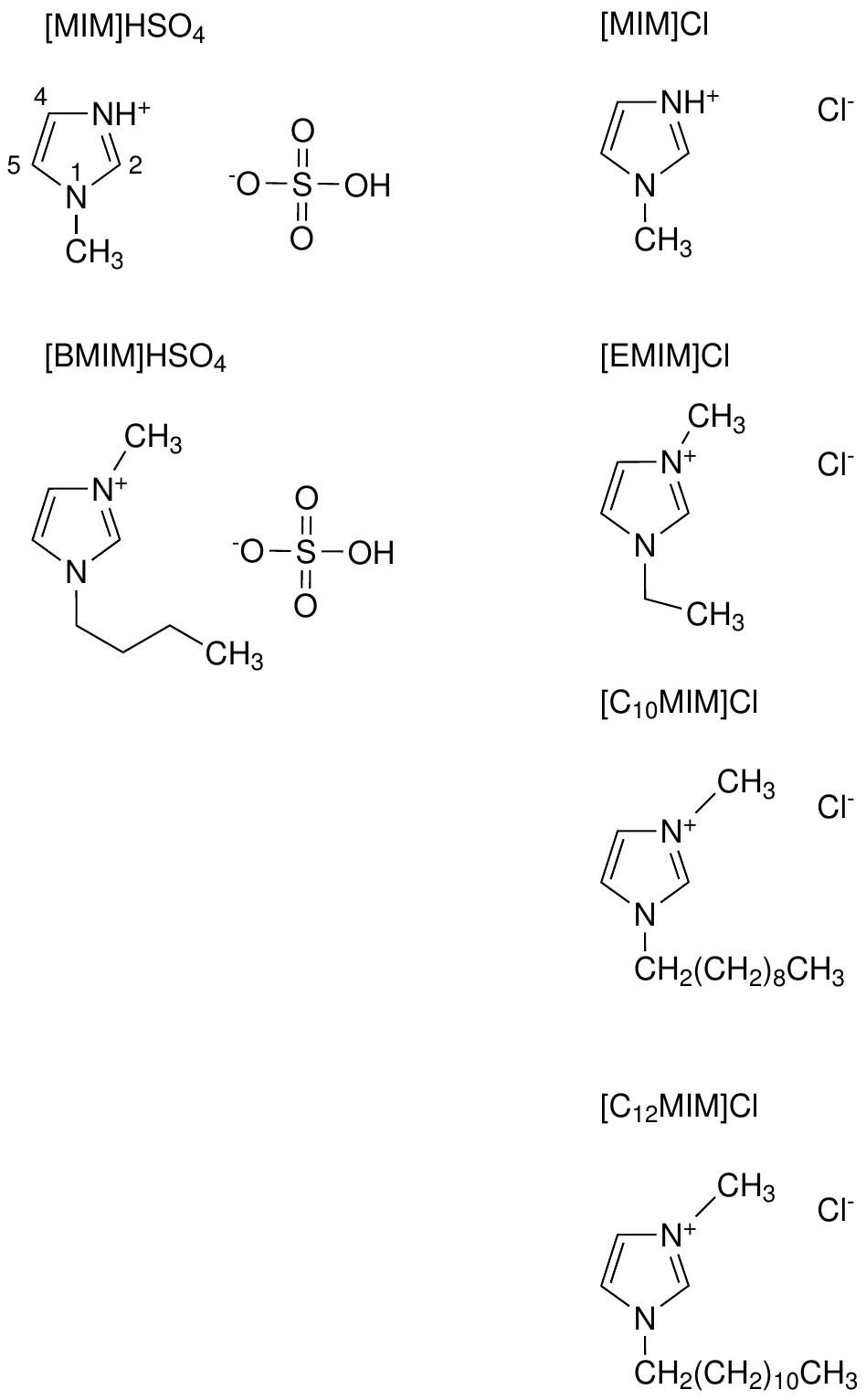}}
\caption{Chemical structure of ILs investigated in this work.} \label{figure1}
\end{figure}

1-methylimidazolium hydrogen sulfate ([MIM]HSO\textsubscript{4}) was purchased from Sigma Aldrich with a purity of 95\%. 1-methylimidazolium chloride ([MIM]Cl), 1-ethyl-3-methylimidazolium chloride \linebreak ([EMIM]Cl), 1-decyl-3-methylimidazolium chloride ([C\textsubscript{10}MIM]Cl), 1-dodecyl-3-methylimidazolium chloride ([C\textsubscript{12}MIM]Cl)  and 1-butyl-3-methylimidazolium hydrogen sulfate ([BMIM]HSO\textsubscript{4}) were purchased from IoLiTec with a purity of 99\%. All the ILs were dried under vacuum (10$^{-3}$ bar) with phosphorus pentoxide for $72$ h in order to remove any possible water contamination before their use. 
High-purity water, deionized through a MilliQ\textsuperscript{TM} water system (>18 M$\Omega$ cm resistivity), was used for all the experiments. IL/H\textsubscript{2}O solutions were prepared in a dry glove box at different molar fraction of IL, $x=n\textsubscript{IL}/(n\textsubscript{IL}+n\textsubscript{H\textsubscript{2}O})$, where $n\textsubscript{IL}$  and $n\textsubscript{H\textsubscript{2}O}$ represent the mole number of IL and water, respectively. The molecular structures of ILs used in this study are reported in figure~\ref{figure1} together with the atom labeling.  
\vspace{6mm}

\vspace{-4mm}
\subsection*{Raman measurements}
UVRR experiments using SR were carried out at the BL10.2-IUVS beamline of Elettra-Sincrotrone Trieste (Italy) \cite{31}. The exciting wavelengths used for collecting Raman spectra were fixed at 235 and 250 nm by adjusting the gap parameters of the undulator and by using a Czerny-Turner monochromator (Acton SP2750, Princeton Instruments) equipped with a 3600 grooves/mm grating to monochromatize the incoming SR. The UVRR spectra were collected in a back-scattered geometry by using a single pass of a Czerny-Turner spectrometer (Trivista 557, Princeton Instruments). The spectral resolution was set to about 6 cm$^{-1}$ in order to have a satisfactorily high signal to noise ratio. The calibration of the spectrometer was standardized using cyclohexane (spectroscopic grade, Sigma Aldrich). The power of the beam on the sample was kept sufficiently low (a few \textmu{}W) in order to avoid photo-damage effects and heating of the sample.

Spontaneous Raman spectra were recorded at 785 nm laser excitation wavelength with a  spectral resolution of 3 cm$^{-1}$. Raman data were collected using the spectrometer MonoVista CRS+ from the company S\&I (Acton spectrometer SP2750 with Princeton Instruments PyLoN CCD camera) operating in a backscattering geometry.
All UVRR and spontaneous Raman measurements have been carried out at room temperature (298 K).

\section{Result and discussion}
\begin{figure}[!b]
\centerline{\includegraphics[width=0.75\textwidth]{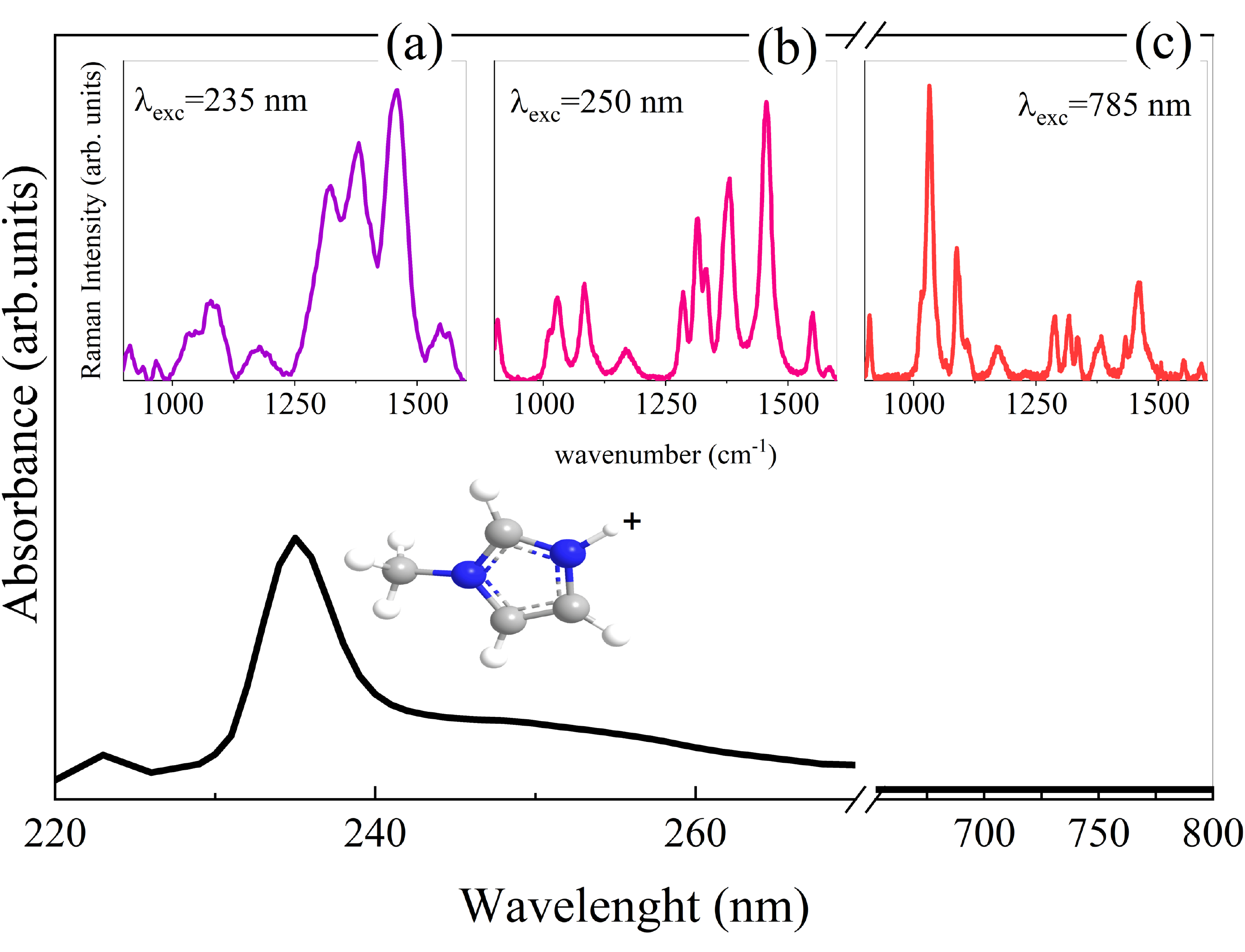}}
\caption{(Colour online) UV-VIS Absorption spectrum of [MIM]HSO\textsubscript{4} in water and Raman spectra collected on pure IL collected at 235 (a), 250 (b) and 785 nm (c) of excitation wavelengths in the spectral range 900--1600 cm$^{-1}$.} \label{figure2}
\end{figure}

The UV-Vis absorption spectrum of [MIM]HSO\textsubscript{4} in water is presented in figure~\ref{figure2}. The curve displays a strong absorption band at about 235 nm and a shoulder at 250 nm going to zero from $\sim$300 nm onwards. Based on the absorption features, Raman spectra of neat [MIM]HSO\textsubscript{4} have been collected with excitation wavelenghts at 785, 250 and 235 nm in order to approach spontaneous, pre-resonance and resonance conditions, respectively. Attention has been focused on the spectral range 1000 to 1600 cm$^{-1}$, although this frequency window is not extensively investigated in literature. The comparison between UV and visible Raman profiles points out an increasing Raman cross section of the vibrational modes between 1250 and 1600 cm$^{-1}$ at low excitation wavelengths. These bands have been assigned to vibrations involving the imidazolium ring \cite{27,28,29,30} in the molecular structure of [MIM]HSO\textsubscript{4} [figure~\ref{figure2}~(a) and (b)]. The enhancement of these peaks in the UVRR spectra was expected due to the  $\pi$-$\pi$\textsuperscript{*} transition of imidazole ring electrons occurring below 300 nm. Conversely, the Raman peak at 1030 cm$^{-1}$ is associated with the HSO$_4^{-}$ symmetric stretching mode of the anion \cite{32}  and it appears particularly prominent in spontaneous Raman scattering [figure~\ref{figure2}~(a)]. These experimental findings suggest that UVRR also allows us to probe slight modifications occurring in the spectral parameters of the Raman vibrations associated with the cations in imidazolium-based ILs, due to the signal enhancement observed in the spectra excited with UV light compared to those obtained with the visible radiation.  	
The comparison between UVRR spectra of [MIM]HSO\textsubscript{4} excited at 235 and 250 nm [figure \ref{figure2}~(a) and (b), respectively] shows similar experimental profiles, although the spectrum in figure~\ref{figure2}~(a) has been collected with a worse resolution with respect to the corresponding one excited at 250 nm. This is due to the self-absorption phenomenon \cite{33, 34} that is particularly strong with respect to the absorption maximum of the sample at 235 nm. In this condition, although Resonance Raman cross sections are generally larger than those of the pre-resonance Raman, the intensity of the Raman signal is dramatically reduced, leading to a significant decrement of the signal-to-noise ratio.  
The continuous tunability of the SR source in the UV range offers a unique opportunity to finely approach such pre-resonance condition that ensures a satisfactory enhancement of the Raman modes associated with the imidazolium ring. However, at the same time it hampers the self-absorption, as observed in the UVRR spectra excited at 250 nm.

\begin{figure}[!b]
\centerline{\includegraphics[width=0.9\textwidth]{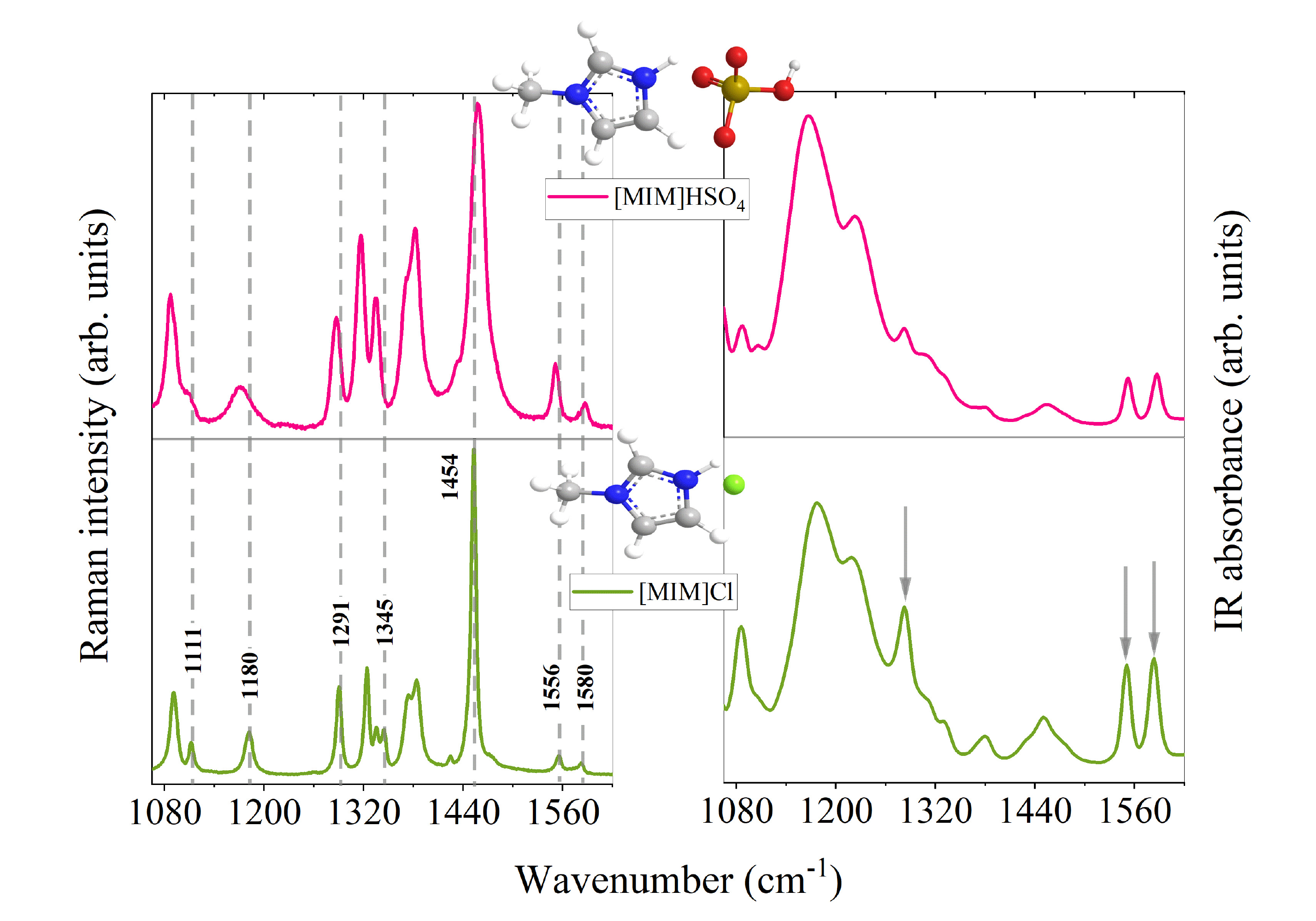}}
\caption{(Colour online) Comparison between UVRR spectra excited at 250 nm (panel on the left) and IR absorption spectra (panel on the right) of neat [MIM]HSO\textsubscript{4} and neat [MIM]Cl.} \label{figure3}
\end{figure}

Figure \ref{figure3} reports the comparison between UVRR and IR spectra of neat [MIM]HSO\textsubscript{4} and [MIM]Cl in the spectral region between 1080 and 1600 cm$^{-1}$, where only the vibrational modes associated with the imidazolium ring can be recognized. The spectra have been collected at room temperature where both ILs are in their crystalline phase. The IR profiles of both [MIM]-based ILs appear to be very similar to each other except for small differences observed for the relative intensities of some IR bands, as highlighted in figure \ref{figure3} by the arrows in the bottom right-hand panel. Conversely, substitution of the anion species in the IL strongly affects the Raman spectra between 1080 and 1600~cm$^{-1}$ (see figure \ref{figure3}, left-hand panel).
In the case of [MIM]Cl with respect to [MIM]HSO\textsubscript{4}, we observe in the Raman spectrum the increment of  a Raman peak at 1111 cm$^{-1}$ (combination of the bending modes involving C(4)H and C(5)H groups on imidazole ring) and the appearance of an additional feature at 1345 cm$^{-1}$, probably related to changes in the molecular organization or in the strength of the interactions involving the imidazolium ring due to the different anion type. The Raman peak centred at 1454 cm$^{-1}$ in the spectrum of [MIM]Cl, assigned to the antisymmetric bending mode of the N-methyl group \cite{35,36}, appears blue-shifted by about 5~cm$^{-1}$ and much broader in the experimental profile of [MIM]HSO\textsubscript{4}. Similarly, also the peaks at 1180 and 1291 cm$^{-1}$ in the spectrum of [MIM]Cl, associated with the bending motions of the CH and NH groups located on the imidazole ring \cite{35}, are found red-shifted by about 5 cm$^{-1}$ and broader in the spectrum of [MIM]HSO\textsubscript{4}. This probably reflects the change in the interaction strength between the anion (Cl$^{-}$ or HSO$_4^{-}$) and the cation in the two different ILs. Finally, also the Raman peaks at 1556 and 1580 cm$^{-1}$ that are associated mainly with vibrations involving the imidazolium ring and the bending modes of the NH groups \cite{29} undergo slight frequency shifts and intensity variation, as can be observed by comparing the UVRR spectrum of [MIM]Cl with the corresponding one of [MIM]HSO\textsubscript{4}. All these findings suggest that UVRR spectra are rich in vibrational signatures of the interactions established between cation and anion that drive the molecular organization of IL domains.

\begin{figure}[!t]
\centerline{\includegraphics[width=0.85\textwidth]{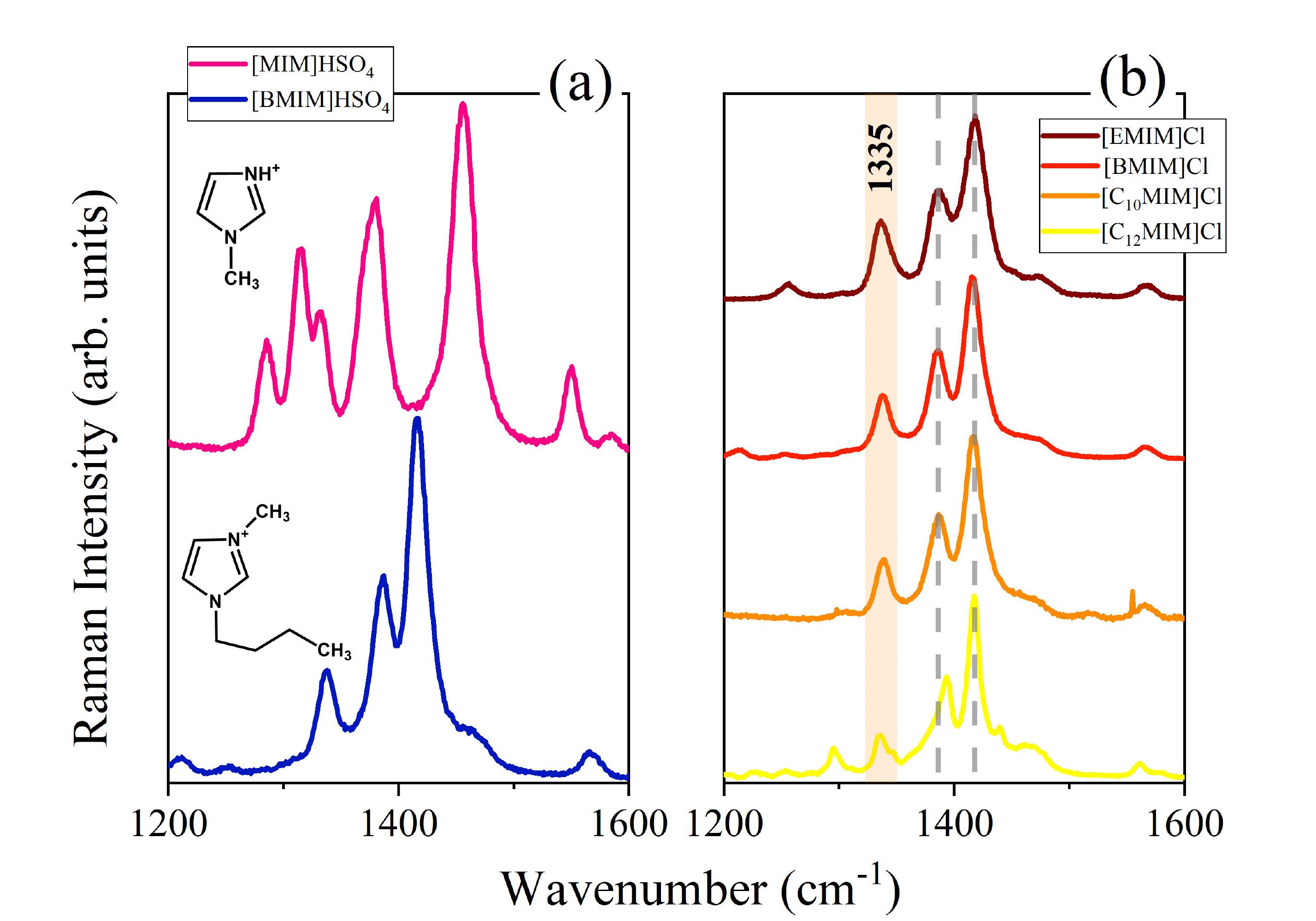}}
\caption{(Colour online) Comparison between the UVRR spectra (excited at 250 nm) of (a) [MIM]HSO\textsubscript{4} and [BMIM]HSO\textsubscript{4} and of (b) [EMIM]Cl, [BMIM]Cl, [C\textsubscript{10}MIM]Cl and [C\textsubscript{12}MIM]Cl.}
	 \label{figure4}
\end{figure}

A further support to this hypothesis can be found by inspection of figure \ref{figure4}. The figure shows UVRR spectra collected on a set of imidazolium-based ILs with N-alkyl chains of different length in the spectral region mainly dominated by the imidazolium ring vibrations. Significant modifications of the spectral profiles associated with the imidazolium vibrational modes seem to be induced by the replacement of the hydrogen atom attached to the imidazolium N with an alkyl chain  [compare the UVRR spectrum of [MIM]HSO\textsubscript{4} to that of [BMIM]HSO\textsubscript{4}  in figure \ref{figure4}~(a)]. The UVRR spectrum of [BMIM]HSO\textsubscript{4} is characterized by three strong bands at about 1335, 1386 and 1415 cm$^{-1}$, assigned to combined vibrational modes mainly involving the imidazolium ring and the CH and NH groups located on the ring \cite{37}. In the same spectral region, the experimental profile of [MIM]HSO\textsubscript{4} appears quite different from that of its analogue [BMIM], confirming that the UVRR modes of imidazolium are strongly sensitive to the chemical structure of the substituent on the ring.

Figure \ref{figure4}~(b) further supports this outcome, giving evidence that the Raman peaks mentioned above can be specifically associated with the  change of substituents on the imidazolium ring in this type of ILs. Although the mode at 1415 cm$^{-1}$ does not exhibit any significant modification in frequency position or intensity when the alkyl-chain is extended from [EMIM] to [C\textsubscript{12}MIM], conversely the peak at 1386~cm$^{-1}$ undergoes a blue-shift of about 5 cm$^{-1}$. At the same time, a progressive reduction of the intensity of the Raman peak at 1335 cm$^{-1}$ can be observed as the length of the alkyl chain on the ring increases. Since this mode corresponds to a combination of ring breathing and stretching motions of N(1)-CH\textsubscript{2} and N(3)-CH\textsubscript{2} for [EMIM]-based ILs \cite{38}, we can propose that the increasing length of the alkyl chain on the imidazolium ring is reflected in a progressive hindering of the stretching vibrations involving N(1)-CH\textsubscript{2}, as shown by the decreasing intensity in figure \ref{figure4}~(b). This suggests that the Raman peak at 1335 cm$^{-1}$ in the UVRR spectra of imidazolium-based ILs could be used as a sensitive spectroscopic marker of the molecular reorganization of cationic domains induced by the progressive elongation of the alkyl substituents at the imidazolium ring. 

\begin{figure}
\centerline{\includegraphics[width=0.85\textwidth]{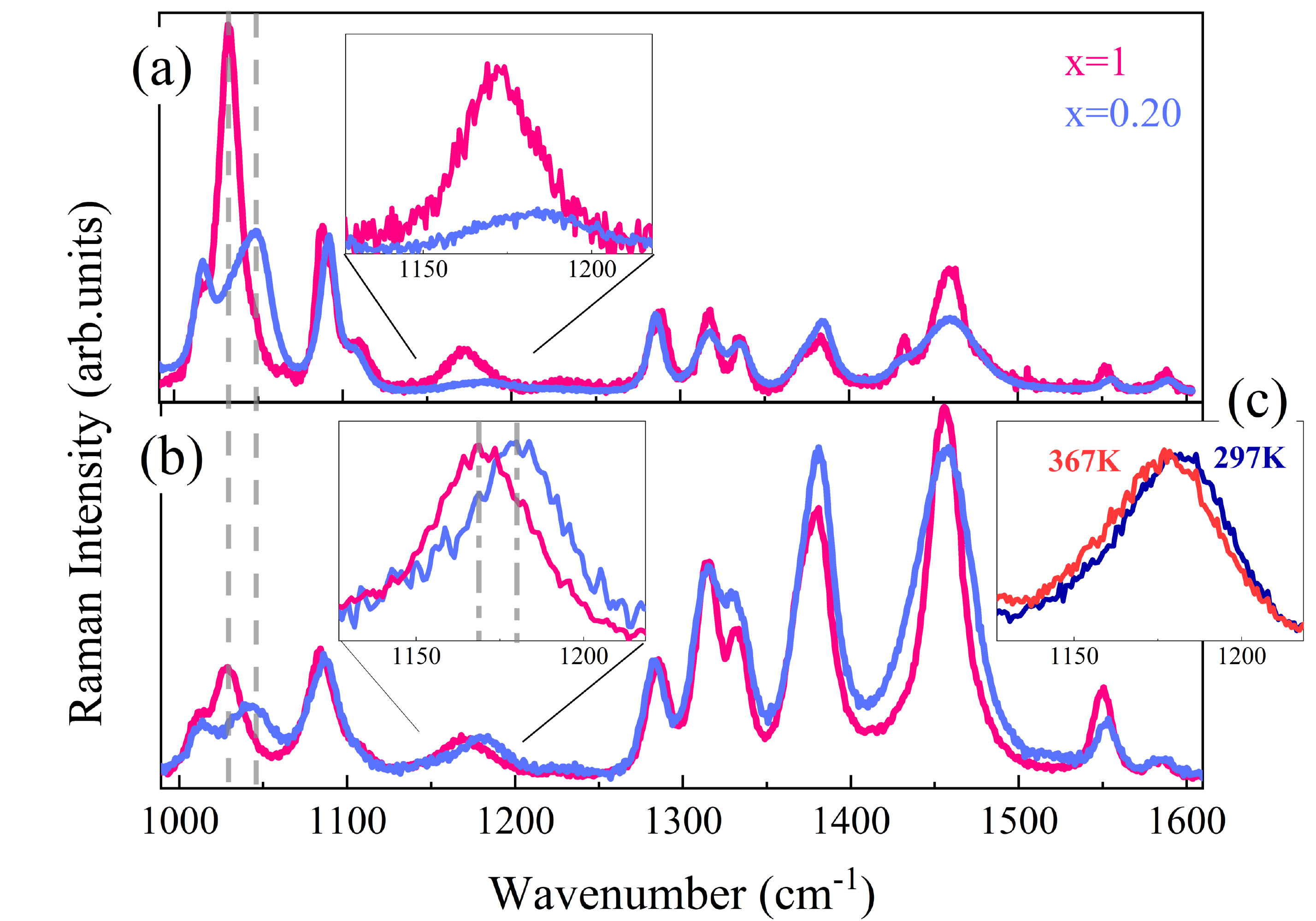}}
\caption{(Colour online) Comparison between the Raman spectra of pure (pink) and hydrated (blue) [MIM]HSO\textsubscript{4} collected at an excitation wavelength of a) 785 nm and b) 250 nm. (c) UVRR spectra of hydrated MIM]HSO\textsubscript{4} (x=0.30) recorded at two different temperatures, i.e., 297 and 367 K collected at 250 nm.} \label{figure5}
\vspace{-2ex}
\end{figure}

Finally, figure \ref{figure5} points out the greater sensitivity of UVRR technique with respect to conventional spontaneous Raman scattering in probing the effect of water on the strength of H-bonds involving cation domains in imidazolium-based ILs. Figure \ref{figure5} shows the comparison between the Raman spectra of pure and water-added [MIM]HSO\textsubscript{4} collected at an excitation wavelength of 785 nm [panel (a)] and 250 nm [panel (b)]. We remark that the analysis of the spectral range considered in figure~\ref{figure5} is usually very difficult to be analyzed by using IR spectroscopy, due to the strong signal of water dominating the spectra in this region also at low water content \cite{26,39}. Raman spectroscopy offers the advantage of exploring also the spectral region between 1000 and 1600 cm$^{-1}$ by looking at the modifications occurring to the Raman signals arising exclusively from ILs also in very high-diluted conditions.
Both the UVRR and spontaneous Raman profiles of figure \ref{figure5} point out that also at this relative low-dilution condition (IL/water molar fraction = 0.2) the Raman peak centred at 1032 cm$^{-1}$ in the neat IL, assigned to the stretching vibration of the anion HSO\textsubscript{4}, shifts toward higher wavenumbers when the water is added to the system. This effect can be explained by considering the solvation effect of water on the anions competing with the electrostatic interactions between cation and anions in the hydrated ILs.   
As concerns the Raman vibrational modes associated only to the molecular motions involving the imidazolium ring, the vibrational spectra appear quite sensitive to the addition of water, especially for the UVRR profiles  [figure \ref{figure5}~(b)]. This is reflected mainly by a slight shift in the frequency position, changes in intensity and broadening of some Raman modes observed in the spectrum of [MIM]HSO\textsubscript{4} as a consequence of hydration. Remarkably, the UVRR spectra of hydrated [MIM]HSO\textsubscript{4} evidence a significant blue-shift of about 10 cm$^{-1}$ of the Raman peak centred in the neat IL at about 1170 cm$^{-1}$ [inset of figure \ref{figure5}~(b)]. This effect is not so clearly detected in the spontaneous Raman spectra of [MIM]HSO\textsubscript{4} [figure \ref{figure5}~(a), inset], probably due to a stronger reduction of the Raman cross section of the modes arising from IL when it is dissolved in water. The main vibrational contribution to the Raman mode at 1170 cm$^{-1}$ has been assigned to a combination of bending modes of the C(2)-H and N(3)-H groups on the imidazolium ring of [MIM]HSO\textsubscript{4} \cite{40}. On the other hand, these two sites have been demonstrated to be responsible for the formation of H-bonds between the cation and anion \cite{41} in the IL. This interpretation is supported by the experimental evidence that the same Raman peak moves  to lower frequencies upon the increasing of thermal motion [figure \ref{figure5}(c)] that usually leads to a weakening of the H-bonds interactions.  
The experimental findings described in figure \ref{figure5}~(b) suggest that for high dilution of [MIM]HSO\textsubscript{4} in water, a complete exchange of water molecules with anions in the formation of hydrogen bonds with cations can be expected and that the strength of this cation-water interaction is probably stronger than the cation-anion one.

\newpage
\section{Conclusions}
This work highlights the advantages of synchrotron-based UV Resonance Raman scattering for the investigation of the local structure and intermolecular interactions in imidazolium-based ILs, both neat and in solution with water. The comparison between UVRR and spontaneous Raman spectra of ILs discussed in the paper underlines an important feature of our approach:  the unique tunability of SR in the UV range permits a selective enhancement of the vibrational signals arising from the imidazolium ring, at the same time avoiding the self-absorption phenomena, through a suitable choice of the pre-resonance conditions. The detailed analysis of UVRR spectra of several imidazolium-based ILs reveals that the Raman modes associated with the cation are sensitive to modifications induced in the local structure of ILs by the substitution of anions, while similar changes are not so efficiently detected by IR spectroscopy in the same spectral region. Similarly, in the UV-resonance spectra we recognized some spectroscopic signals as a marker of the molecular reorganization of cationic domains induced by the progressively larger alkyl chain substituents at the imidazolium ring. Finally, the UVRR technique turned out to be particularly informative in probing the effect of water on the strength of cation-anion interactions in imidazolium-based IL and water solutions. We focused in particular on the spectral region between 1000 and 1600 cm$^{-1}$, a spectral range usually not accessible by IR or spontaneous Raman scattering due to the interfering signal arising from water, especially in high diluted conditions. This experimental approach allowed us to recognize a specific vibrational signal associated with the CH and NH groups located on the imidazolium ring acting as H-bond acceptors. The significant frequency shifts observed for this signal only in UVRR spectra for hydrated [MIM]HSO\textsubscript{4} suggest an increase in the strength of the H-bond interactions between cation and anions upon adding the water.
All these results demonstrate the great potentiality of UV Raman spectroscopy to retrieve the information on the intermolecular interactions in IL-water solutions, besides the counterpart obtained by employing the spontaneous Raman scattering technique on these systems.

\section*{Acknowledgements}
The authors acknowledge the CERIC-ERIC Consortium for the access to experimental facilities and financial support. B.C. gratefully acknowledges Dr. Paolo Zucchiati and Dr. Barbara Giabbai for the support during FTIR and UV/Vis absorbance measurements, respectively.

\ukrainianpart

\title{Синхротронне УФ резонансне раманівське розсіювання для досліджень водних розчинів іонних рідин}
\author{К. Боттарі\refaddr{label1,label2}, Б. Россі\refaddr{label1}, A. Meлє\refaddr{label3,label4}, A. Дамін\refaddr{label5}, С. Бордіга\refaddr{label5}, M. Mуссо\refaddr{label6}, A. Гессіні\refaddr{label1}, К.~Maшіовеккіо\refaddr{label1}
}
\addresses{
	\addr{label1} Elettra-Сихротрон Трієст, 34149 Трієст, Італія
	\addr{label2} Фізичний факультет, Університет Трієсту, 34127 Трієст, Італія
	\addr{label3} Факультет хімії, матеріалів та хімічної інженерії ``Дж. Натта'', Політехніка Мілану, 20133 Miлан, Італія
	\addr{label4} CNR-ICRM, 20131 Мілан, Італія
	\addr{label5} Факультет хімії, Центр NIS та Центр INSTM, Університет Турину, 10135 Турин, Італія
	\addr{label6} Факультет хімії та фізики матеріалів, Університет Зальцбурга, Зальцбург, Австрія
}

\makeukrtitle

\begin{abstract}
	Ця робота показує, що об'ємні іонні рідини (ІР) та їх водні розчини можуть зручно бути дослідженими резонансною раманівською (РР)
	спектроскопією на основі синхротронного ультрафіолету (УФ). Представлено та обговорено головні переваги цієї техніки для дослідження 
	локальної структури та міжмолекулярних взаємодій у ІР на основі імідазоліуму. Унікальне підстроювання синхротронного джерела дозволяє
	селективно підсилити у раманівських спектрах вібраційні сигнали з кілець імідазоліуму. Такі сигнали показали добру чутливість до модифікацій
	індукованих в локальній структурі ІР при  i) заміні аніону та ii)~заміні прогресивно довгого алкілового ланцюга на кільце імідазоліуму. 
	Однак, деякі сигнали УФРР є особливо інформативні стосовно ефекту індукованого додаванням води на силу катіон-аніонного водневого 
	зв'язку в розчинах ІР-вода. Усі ці результати підтверджують потенціал УФРР для отримання інформації про міжмолекулярні взаємодії 
	у водних розчинах ІР, до того ж аналогічне було отримано для цих систем технікою спонтанного раманівського розсіювання.

	\keywords іонна рідина, УФ резонансне раманівське розсіювання, міжмолекулярна взаємодія, водневий зв'язок
	
\end{abstract}

\end{document}